\documentclass[a4paper,fleqn]{cas-sc}
\pdfoutput=1
\usepackage{epsfig,multicol}
\usepackage{dcolumn}
\usepackage{amssymb,amsmath}
\usepackage{float}
\usepackage[square,sort,comma,numbers]{natbib}
\usepackage{graphicx}
\usepackage[usenames,dvipsnames]{pstricks}
\usepackage{braket}
\usepackage{color}
\usepackage{textcase}
\usepackage{subfig}
\usepackage{epstopdf}
\usepackage{booktabs}
\usepackage{lipsum}
\usepackage{caption}
\usepackage{lineno}
\graphicspath{{fig/}}
\begin{document}
\let\WriteBookmarks\relax
\def\floatpagepagefraction{1}
\def\textpagefraction{.001}
\shorttitle{On LGRB progenitors: an approach from thermally-produced neutrinos }
\shortauthors{G. Morales and N. Fraija}
\title[mode=title]{On LGRB progenitors: an approach from thermally-produced neutrinos }
\author[1]{G. Morales}[orcid=0000-0002-2976-4739] 
\cormark[1]
\ead{gmorales@astro.unam.mx}
\address[1]{Instituto de Astronom\' ia, Universidad Nacional Aut\'onoma de M\'exico, Circuito Exterior, C.U., A. Postal 70-264, 04510, Ciudad de M\'exico, Mexico}
\cortext[cor1]{Corresponding author}
\author[1]{N. Fraija}
\ead{nifraija@astro.unam.mx}

\begin{abstract}
Gamma-ray bursts (GRB) are the most intense electromagnetic (EM) sources in the Universe. Long GRB (LGRB) correspond to those events with a typical prompt emission of more than a few seconds. It is generally assumed that they are originated after an implosion of a very massive star within a central compact object engine that can be either a black hole (BH) or a rapidly-spinning highly-magnetized neutron star (NS). Nevertheless, one of the most challenging aspects of defining a unique model is that the progenitor remains initially hidden for direct EM observation. In this work, we investigate the evolution of thermally-produced neutrino properties in both GRB progenitors to provide an alternative solution. We consider the characteristics of both progenitors and the fireball scenario to calculate the oscillation probabilities within a three-flavor admixture regime. Then we obtain the expected neutrino ratio and we also estimate the number of events from these sources that could be detected in the future Hyper-Kamiokande (Hyper-K) detector, considering a sample of previously observed GRB with remarkably signs of being magnetar-produced. Our findings indicate that examining the predicted neutrino rates result in an additional mechanism to determine the type of progenitor associated with these events. This is especially useful when, for instance, we cannot directly observe an electromagnetic counterpart, such as so-called "failed" GRB with hidden jets, or when light curve analysis is inconclusive. 

\date{\today}
\end{abstract}
\maketitle
\section{Introduction}

Observations over several years revealed that GRB have a bimodal distribution in terms of prompt emission time, with an explicit separation around two seconds \citep{1993ApJ...413L.101K}. This bimodality suggests the existence of two different types of progenitors: short Gamma-ray bursts (sGRB) and long Gamma-ray bursts. Short timelines (on the order of milliseconds) in the case of sGRB imply a notion of a progenitor based on the merging of two compact objects, such as neutron star - neutron star  or black hole - neutron star  \citep{1992ApJ...392L...9D, 1992Natur.357..472U, 1994MNRAS.270..480T,lee2007progenitors,berger2014short}. LGRB progenitors, on the other hand, are connected with the core-collapse (CC) of a massive star, resulting in an Ic-type supernova due to their unique placement in low-metallicity host galaxies with active star formation \citep {woosley1993gamma, paczynski1998gamma, macfadyen1999collapsars, 1998Natur.395..670G, 1999Natur.401..453B, 2006ARA&A..44..507W, lun15}.  Massive star collapse can occur in one of two ways. The most common involves the creation of a hyper-accreting stellar-mass black hole \citep{1993ApJ...405..273W,nar01} from which a relativistic jet is launched by $\nu\bar{\nu}$- annihilation processes \citep{ruf97,pop99} or  the Blanford-Znajek (BZ) mechanism \citep{bla77}, whereas in the second scenario, a rapidly spinning, and strongly magnetized neutron star ("magnetar") is formed  with enough rotational energy to avoid gravitational collapse \citep{1992Natur.357..472U, 1998A&A...333L..87D, 2000ApJ...537..810W, 2007MNRAS.380.1541B, 2011MNRAS.413.2031M}. In either case, a spinning disk of long-lived debris is left near the compact object. Because the temperature exceeds the rate of $e^\pm$ pair production, nuclei are photo-disintegrated, and the plasma formed at the base of the progenitor so-called fireball is predominantly made up of free pairs, gamma-ray photons, and baryons \citep{pir99}.\\

During the afterglow episode, the X-ray light curves, apart from exhibiting a flare, have a typical profile consisting of four different power-law (PL) functions $\propto t^{-\alpha_{\rm x}}$ \citep{2006ApJ...642..354Z};  the initial steep decay, the ``plateau" phase,  the normal decay and the post-jet break phase.  In some cases, the ``plateau" phase is followed by  an abrupt steep decay with a falling slope steeper than $\alpha_{\rm x}\gtrsim3$ \citep{2007ApJ...665..599T,2007ApJ...670..565L,2010MNRAS.402..705L,2012A&A...539A...3B} instead of the normal decay phase. Currently, the most accepted model to explain this late activity is the existence of a long-lasting BH/magnetar central engine that continues supplying energy to the system for a time longer than the duration of the prompt emission  (up to $\sim 10^4-10^5$ s) through the dissipation of the leptonic wind and the surrounding medium. The flares observed may also be an evidence of magnetic readjustments within the magnetar. Several authors have statistically identified these attributes in a large sample of past GRB observations, so this central engine model has gained much importance \citep{ber13,lu14,li18,hou21}. On the other hand, the abrupt steep decay disfavors an external forward-shock scenario and suggests that the EM emission stopped abruptly. This atypical signature has been associated with the collapse of a rapidly spin-down magnetar into a black hole when it loses centrifugal support  \citep{2007ApJ...665..599T, 2014ApJ...780L..21Z, 2011MNRAS.413.2031M}.  \\

While these central engine models have become more robust in recent years, and much work has been done in this field, many unknowns still remain unclear. In this regard, distinguishing between the two inner engine models that produce GRB is essential for understanding not just these extremely energetic events, but also their associated progenitors and primary emission mechanisms. However, this task becomes more complicated when we consider that the photon opacity within the core during the initial phase is so high that we cannot directly observe the internal processes that cause these bursts. As a result, drawing conclusions based solely on light curve analysis is problematic. In this context, neutrinos turn out to be a valuable detection channel because they are weakly interacting particles and have a smaller cross-section than photons. It is also estimated that $\sim99\%$ of the gravitational binding energy is released as neutrinos throughout the core collapse, serving as an effective cooling mechanism for the system \citep{woo93,hal96}. This implies that an energetic burst located close enough would produce a significant neutrino flux that could be detected in future Megaton neutrino telescopes, as some authors propose \citep{liu16,fra16,kyu18,morales2021differentiating}. Neutrinos from astronomical sources have previously been observed, but  only the multiple MeV-neutrinos from SN1987A were the first detected particles associated with a single point source \citep{SN1987A,bio91}. This multi-messenger (photons and neutrinos) scenario signaled the beginning of a new era of astronomical observation.\\

Although many authors have studied the conversion of neutrino flavors attributed to interactions with surrounding matter via the Mikheyev-Smirnov-Wolfenstein \textit{(MSW)} effect \citep{zhu16,fre17,wu17,morales2021winds} and the neutrino propagation inside a fireball during the formation of a GRB \citep{dol92,bah00,koe05,sah09}, even considering a strong magnetic contribution \citep{fra14,pal15}. None of them have considered neutrino detectability as a mechanism to differentiate between both inner engine models.
Therefore, in this work, we incorporate the multi-messenger nature of GRB to address some of these open questions by studying thermal neutrino flavor evolution. We focus on those neutrinos with energies lying in the MeV range, propagating across a magnetized fireball in both central engine models. \\

This paper is organized as follows. In Section \ref{sec:neutrino} we provide the neutrino effective potential in a magnetized fireball and present the neutrino formalism used during propagation in this medium. A brief review of both central engine models within LGRB, as well as a compendium of some LGRB that have shown signs of being magnetar originated, is shown in Section \ref{sec_lgrb}. In Section \ref{sec:results}, we compute the three-flavor oscillation probabilities and the flavor neutrino ratio for both  BH/magnetar scenarios. Additionally, we show the number of events that Hyper-K could potentially detect in the coming years. Lastly we summarize our Conclusions in Section \ref{sec:conclusion}. Unless otherwise specified, we use the system of natural units where $(k=c=\hbar=1)$ as well as, the convention of $Q=Q/10^x$ in cgs units for the remainder of this article.
We have also considered an Einstein-deSitter Universe with parameters $h=0.673$, $\Omega_\Lambda=0.685$, and $\Omega_m=0.315$ \citep{2018arXiv180706209P}, where  $\Omega_\Lambda$ represents the dark energy density of the $\Lambda$CDM Universe and $\Omega_m$ denotes the pressureless matter density.

\section{Neutrino phenomenology}\label{sec:neutrino}

Neutrinos are particles which may interact weakly with background particles via neutral current (NC) and charge current (CC) interactions. They are classified into  three flavors associated with the three different leptonic families. 
Neutrinos oscillate with one another, and their transitions can be calculated analytically in the vacuum or through a material influenced by an effective potential equivalent to the medium's refractive index \citep{mikheyev1986yad}. This additional potential increases the neutrino effective mass and modified the flavor eigenstates. The time evolution of neutrinos is obtained by solving the Schr\"odinger equation $ i\ket{{\dot{\nu}_\alpha (t)}}=\mathcal{H}_f\ket{\nu_\alpha(t)}$ for the neutrino state within the density matrix formalism, assuming a plane wave approximation, whose solution is
  \begin{equation}\label{eq:Sch_solution}
      \ket{\nu_\alpha(t)}=e^{-i\mathcal{H}_f t}\ket{\nu_\alpha(t)}=\mathcal{U}_f(t)\ket{\nu_\alpha(t)}\ ,
  \end{equation}
 where $\mathcal{U}_f$  represents the temporal evolution operator and $\mathcal{H}_f$  denotes the associated Hamiltonian which incorporates the system's characteristics.  Both operators are reflected in the flavor basis, as indicated by the subindex $f$. The transition probability between neutrino flavor $\alpha$ and neutrino flavor $\beta$ is then determined by the square of the probability amplitude between neutrino states $\alpha$ and $\beta$ 
 
 \begin{equation}\label{eq:Proba_m}
 P_{\alpha\beta}\equiv A_{\alpha\beta}^2=|\braket{\nu_\beta|\mathcal{U}_f|\nu_\alpha}|^2=\left(\sum_\alpha\sum_\beta e^{-i(\mathcal{H}_{f,\beta}-\mathcal{H}_{f,\alpha})\ t}\braket{\nu_\beta|\nu_\alpha} \right)^2 .
 \end{equation}
 
In practice, determining the probabilities is more accessible in a linearly independent mass basis with a diagonal Hamiltonian, which requires an additional basis transformation  $\mathcal{U}_f=e^{-i\mathcal{H}_{f,\rm mat}\ L}$ with $\mathcal{H}_{f,\rm mat}=U\ \mathcal{H}_{m,\rm mat}\ U^{-1}$, where $U$ denotes the unitary Pontecorvo-Maki-Nakagawa-Sakata (PMNS) matrix \citep{gro74}, and where we have also taken into account the approximation that neutrinos propagate at relativistic velocities $t\sim L$.  In fact, the probabilities of vacuum oscillation can be determined analytically for this case.
However, when matter effects are considered into the Hamiltonian, the calculations quickly become complicated, and the oscillation probabilities can only be calculated numerically for specific conditions.
As a result, this study uses the general-purpose \texttt{NOPE} code \citep{bus19} to handle neutrino propagation in both internal engine models for a three-flavor admixture scenario. It is worth mentioning that vacuum oscillations are suppressed once neutrinos leave the source and travel to Earth since they have already been polarized into incoherent mass eigenstates \citep{lun01,rom15,smi05,kne08}.  Therefore, this effect will no longer be considered in this work. \\

\subsection{Neutrino potentials within a magnetized fireball}
In order to incorporate the medium effect, in this paper, we employ Fraija's neutrino potential for a magnetic fireball within two regimes \citep{fra14}. The first scenario corresponds to a strong magnetic field above the critical magnetic field $(\Omega_B>1)$ where $\Omega_B\equiv B/B_c$ , represents the dimensionless parameter of the magnetic field as a function of the Landau's critical magnetic field of the electrons defined as $B_c=m_ec^2/2\mu_e=m_e^2c^3/e\hbar\sim4.414\times 10^{13}\, {\rm\ G}$.  The second scenario describes a mild magnetic field with $\Omega_B<1$.
\subsection*{Strong $\vec{B}$ limit $(\Omega_B>1)$}

\begin{eqnarray}\label{eq:veffs}
V_{\rm eff,s}=\frac{\sqrt2\,G_F\,m_e^3 B}{\pi^2\,B_c}\biggr[\sum^{\infty}_{l=0}(-1)^l\sinh\alpha_l   \left[F_s-G_s\cos\varphi \right]\nonumber\\
-4\frac{m^2_e}{m^2_W}\,\frac{E_\nu}{m_e}\sum^\infty_{l=0}(-1)^l\cosh\alpha_l  \left[J_s-H_s\cos\varphi \right]  \biggr]\,.\cr
 \end{eqnarray}

\subsection*{Mild $\vec{B}$ limit $(\Omega_B<1)$}

\begin{eqnarray}\label{eq:veffw}
V_{\rm eff,m}=\frac{\sqrt2\,G_F\,m_e^3 B}{\pi^2\,B_c}\biggr[\sum^{\infty}_{l=0}(-1)^l\sinh\alpha_l  \left[F_m-G_m\cos\varphi \right]\nonumber\\
-4\frac{m^2_e}{m^2_W}\,\frac{E_\nu}{m_e}\sum^\infty_{l=0}(-1)^l\cosh\alpha_l \left[J_m-H_m\cos\varphi \right]\,,\cr 
\end{eqnarray}
where $m_e$ is the electron mass, $\alpha_l=(l+1)\mu/T$  with $\mu$ and $T$ the chemical potential and temperature, respectively, $E_\nu$ is the neutrino energy, { $\varphi$ is derived from the neutrino dispersion relation and represents the angle between the neutrino momentum, and the direction of the magnetic field.} The functions F$_s$, G$_s$, J$_s$, H$_s$, F$_m$, G$_m$, J$_m$, and H$_m$ are  described in the Appendix.

\subsection{Neutrino global fits}
We show in Table \ref{table} the most recent global fits for three-flavor neutrino oscillations in normal-ordering (NO) and inverted-ordering (IO) configurations \citep{sal21}.
\begin{table}[htbp!]\centering
\catcode`?=\active \def?{\hphantom{0}}
\begin{tabular}{|c|cc|}
\hline
Parameter & Best fit $\pm$ $1\sigma$ (NO) & \hphantom{x} Best fit $\pm$ $1\sigma$ (IO)  \hphantom{x} 
\\ \hline
$\Delta m^2_{21} [10^{-5}$eV$^2${]}        & $7.50^{+0.22}_{-0.20}$                     & $7.50^{+0.22}_{-0.20}$                     \\
$\Delta m^2_{31}| [10^{-3}$eV$^2${]}      & $2.55^{+0.02}_{-0.03}$                     & $2.45^{+0.02}_{-0.03}$                     \\
$\sin^2\theta_{12} / 10^{-1}$              & $3.18\pm0.16$                              & $3.18\pm0.16$                              \\
$\theta_{12}\ \rm(degree)$                     & $34.3\pm1.0$                               & $34.3\pm1.0$                               \\
$\sin^2\theta_{23} / 10^{-1}$              & $5.74\pm0.14$                              & $5.78^{+0.10}_{-0.17}$                     \\
$\theta_{23} \ \rm(degree)$                     & $49.26\pm0.79$                             & $49.46^{+0.60}_{-0.97}$                    \\
$\sin^2\theta_{13} / 10^{-2}$              & $2.200^{+0.069}_{-0.062}$                  & $2.225^{+0.064}_{-0.070}$                  \\ 
$\theta_{13}\ \rm(degree)$                     & $8.53^{+0.13}_{-0.12}$                            & $8.58^{+0.12}_{-0.14}$               \\ \midrule
\end{tabular}
\caption{Updated global three-neutrino oscillation parameters summarized by \citep{sal21}.}
\label{table}
\end{table}

\subsection{Resonance energies}
In order to account for resonance effects within the source, we compute the two corresponding resonance energies  in the medium in a three-flavor admixture scenario. These are defined as follows: \citep{raz10}
\begin{equation}
    \label{eq:res_energy}
    E_{\rm res}^{L}\approx\frac{\Delta m_{21}^2}{2V_{\rm eff}}\cos2\theta_{12},\ \ \      E_{\rm res}^{H}\approx\frac{\Delta m_{31}^2}{2V_{\rm eff}}\cos2\theta_{13}\,.
\end{equation}

These energies describe well-defined regions where neutrino flavor conversion within a material medium becomes dominant, owing to the increased contribution to the Hamiltonian. $E_\nu<E_{\rm res}^{L}$ depicts the domain of vacuum transitions while $E_\nu>E_{\rm res}^{L}$ indicates the region dominated by matter effects, which remains still within the adiabatic regime.

\subsection{Neutrino ratio parametrization}
Because terrestrial detectors cannot measure oscillation probabilities, they must rely on physically quantifiable variables. So the probability matrix must be expressed in terms of the expected neutrino flavor ratio. {This is accomplished by incorporating the neutrino flux before and after oscillations take place $\Phi= P_{\alpha\beta}\ \Phi^0$, where $P_{\alpha\beta}$ denotes the probability matrix between the initial $\Phi^0=(\Phi_e^0,\Phi_\mu^0,\Phi_\tau^0)$ and final $\Phi=(\Phi_e,\Phi_\mu,\Phi_\tau)$ neutrino fluxes}. Although there are many ways to do this, for simplicity, we will use the parameterization proposed by \citep{palladino2015}. 

\begin{equation}
    \label{eq:xi_n}
    \xi_n^0=\Phi_n^0/\sum_n \Phi_n^0\ \hspace{2cm}  \xi_n=\Phi_n/\sum_n \Phi_n ,
 \end{equation}
 where  $\xi_n$ is defined as the fraction of neutrinos with a defined flavor $n$. Thus, if the initial neutrino fraction $\xi_n^0\equiv (\xi_e,\xi_\mu,\xi_\tau)^{0}\equiv(f,g,h)$ is known, the neutrino rate after propagation can be parametrized as 
\begin{align}
    \label{eq:xi}
    \xi_e=&\frac{1}{3}+(2-3g-3h)P_0+(g-h)P_1\ ,\nonumber\\
     \xi_\mu=&\frac{1}{3}+\frac{1}{2}(-2+3g+3h)P_0+(1-2g-h)P_1+(g-h)P_2\ ,\nonumber\\
     \xi_\tau=&\frac{1}{3}+\frac{1}{2}(-2+3g+3h)P_0+(-1+g+2h)P_1-(g-h)P_2\ ,\nonumber\\
\end{align}

here $P_0$ $P_1$ and $P_2$  is expressed in terms of the probabilities as follows: 

\begin{align}
    \label{eq:P_i_Palladino}
    P_0=&\dfrac{P_{ee}-\frac{1}{3}}{2}\ ,\nonumber\\
    P_1=&\dfrac{P_{e\mu}-P_{e\tau}}{2}\ ,\\
    P_2=&\dfrac{P_{\mu\mu}+P_{\tau\tau}-2P_{\mu\tau}   }{4}\ .\nonumber
\end{align}

\subsection{Detection}

Future detectors are expected to detect thermal neutrinos from extragalactic sources, as has already been done with the detection of multi MeV-neutrinos from the core-collapse of SN1987 from the Large Magellanic Cloud \citep{SN1987A,liu16}.
Because of the small cross-section of these thermal neutrinos, it is extremely difficult to detect them directly. Therefore, we must rely on {Cherenkov} radiation water detectors to reconstruct the trace of charged leptons produced during inverse $\beta$-decays interactions with the incident neutrinos and water target protons. Among all of the neutrino telescopes planned to build  in the near future, the Hyper-K detector is the most promising one because it will have a greater volume and fiducial mass than its predecessors.

\subsubsection{Hyper-Kamiokande}

The Hyper-Kamiokande detector will be a third-generation {Cherenkov} detector (replacing the present Super-Kamiokande detector), reportedly placed in a Japanese mine. It is expected to begin operations around the second half of this decade. The initial concept depicts two nearly-cylindrical tanks holding (0.56) million metric tons of ultra-pure water. This detector will have 99,000 photomultipliers (PMTs) evenly distributed among the tanks. In addition to Super-Kamiokande's responsibilities, Hyper-K will detect neutrinos from both terrestrial and extraterrestrial sources, as well as, conduct particle physics research on topics, such as, CP violation in the leptonic sector, proton decay, and neutrino oscillation events. \citep{hk18}.\\

\subsection{Number of neutrino expected events}

The number of events that can be observed on Hyper-K can be estimated. as\\
%
%
\begin{equation}
N_{\rm ev}=V_{\rm det} N_A\,  \rho_N  \int_{E'} \int_{t'} \sigma^{\bar{\nu}_ep}_{cc} \left(\frac{dN}{dE}\right)\,dt\ dE\,,
\end{equation}
\noindent where $V_{\rm det}$ is the effective volumen of water, $N_A=6.022\times 10^{23}$ g$^{-1}$ is the Avogadro's number, $\rho_N=(M_{\rm fiducial}/ V_{\rm det})=2/18\, {\rm g\, cm^{-3}}$ is the nucleons density in water, $ \sigma^{\bar{\nu}_ep}_{cc}\simeq 9\times 10^{-44}\,E^2_{\bar{\nu}_e}/{\rm MeV}^2$  is the neutrino cross-section  \citep{1989neas.book.....B},   $dt$ is the neutrino emission time and $dN/dE$ is the neutrino spectrum, so the number of events can be approximated as
\begin{equation} \label{eq:Nev}
    N_{\rm ev}\simeq \dfrac{N_A \rho_N\sigma_{CC}^{\bar{\nu}_ep}}{4\pi d_z^2\braket{E_{\bar{\nu}_e}}}V_{\rm det}\  E_{T,\bar{\nu}_e}\,,
\end{equation}

{where $d_z$ is the distance from neutrino production to Earth, $\braket{E_{\bar{\nu}_e}}$ is the average energy of electron antineutrino and $E_{T,\bar{\nu}_e}=\int L_{\bar{\nu}_e} dt$  is the total neutrino energy emitted \citep{2004mnpa.book.....M, 2014MNRAS.442..239F}. The relation $L_\nu=4\pi d_z^2 F_\nu\braket{E}_\nu=4\pi d_z^2 F_\nu E_\nu^2(dN/dE_\nu)$ was taken into account and we have considered that  neutrino flux luminosity is correlated with the total photon flux as \citep{hal07}
\begin{equation}
    \displaystyle\int\dfrac{dN_\nu}{dE_\nu}E_\nu dE_\nu \propto \int\dfrac{dN_\gamma}{dE_\gamma}E_\gamma dE_\gamma\ ,
\label{eq:Lg_Lnu}
\end{equation}

 some authors even suggest that a large total amount of the energy released during these events is emitted in the form of neutrinos with the proportion
 $L_\nu=10^2L_\gamma$ during the prompt emission \citep{hal96,lat21}.}

\section{LGRB central engine review}\label{sec_lgrb}
It is believed that LGRB are formed as a result of massive star collapses and even though they have been studied for more than a half-century, much remains unknown about the dynamics of their progenitors.
Several theories have been proposed to describe the possible central engines during this time period. So far, the most successful models are those that can describe the following characteristics observed during the prompt-emission and afterglow phases: i)
the progenitors, in particular, must have a large energy reservoir capable of launching an ultra-relativistic outflow suitable for a GRB ($\sim10^{49}-10^{55}$ erg), ii) the source must last long enough with remarkable intermittency to match the variability of the observed X-ray light curves \citep{zha06,troja2017a}, and iii)
a large toroidal magnetic field is also expected in some cases, resulting in the formation of a magnetically-dominated outflow \citep{zha09}. Within this framework, there are two promising progenitor models that meet these criteria. On the one hand, there is a black hole-disk system, and on the other, a millisecond magnetar, whose fast rotation required by the GRB central engine prevents its total collapse to a BH \citep{bur07,des08}. A summary of these candidates' characteristics is provided below.

\subsection{Black hole--accretion disk within the collapsar model}
The collapsar model describes how a very massive star (typically larger than 30 M$_\odot$) \citep{pod04} loses hydrostatic equilibrium and succumbs to gravitational collapse in the core during its main-sequence evolution \citep{woo93}. During this phase, the star loses its external envelopes, which had a lot of angular momentum at the beginning. As a result, all the material in the star does not collapse directly into a black hole but rather forms a system consisting of a rotating black hole and an accretion disk. The gravitational potential energy contained is subsequently transformed into kinetic energy in the form of an ultra-relativistic jet along the rotating axis, while the accretion of surrounding material fuels the jet by either electrodynamic \citep{bla77} or neutrino-antineutrino annihilation processes \citep{ruf97,pop99,che07}. Finally, the newly formed winds are injected from the torus debris via the Poynting flux before being converted into gamma rays.\\

The advantages of this model include that it can produce the high energy that has been recorded for some GRB. Still, it is difficult to explain the prolonged activity of the central engine that is sometimes found in certain X-ray light curves. Similarly, It is estimated that the inhomogeneous accretion of the BH-disk system tends to generate a steep decay rather than a smooth plateau \citep{du20}. The late accretion rate in this scenario follows a fall-back rate of $t^{-5/3}$, and the total energy is then determined either by  the total accreted mass  $E_{tot}=\eta m_{acc}c^2$ \citep{lev16} or by the spin energy of the BH. In the first case, assuming an energy conversion efficiency factor $\eta=0.1$ and an accretion mass of $\sim 10\ M_\odot$, the total energy is about $E_{tot}\sim2\times10^{54}$ erg. In the later case, the associated rotational energy of the black hole is also close to this value $E_{rot}\sim 2\times10^{54}$ erg $f_{rot}(a_\bullet)(M/M_\bullet)$, where $f_{rot}(a_\bullet)$ represents a function in terms of the BH spin parameter $a_\bullet$\citep{li18}. 
These values match  the order of the most energetic GRB detected. Lastly, the associated magnetic field of the BH is estimated to be $\sim10^{10}-10^{12}$ G \citep{tsu18,mor21}, depending on the energy extraction mechanism considered.
In any case, these typical  values are  less than the critical magnetic field ($\Omega_B<1$). 

\subsection{The millisecond magnetar model}
In this scenario, a massive star first collapses to form a highly magnetized neutron star, converting all of the star's gravitational potential energy into rotational energy with a rotational period on the millisecond scale. When the magnetar is born, a plethora of neutrinos is produced via pair-annihilation eventually resulting in a fireball made primarily of leptons and photons. Because of the little amount of baryonic compound, the fireball expands relativistically. The neutrino outflow acts as an effective cooling mechanism for the system, dragging and heating the surrounding material in the so-called neutrino-driven winds. As a result, more neutrinos are created through the thermal interactions between baryons present in the winds and fireball electrons (cf. Section \ref{subsec:fireball})
 with  an average energy ranging about 8 and 25 MeV \citep{rosII}. For a magnetar-like magnetic field, this wind is accelerated by the dynamo mechanism and is more energetic than the first one, so the neutrino outflow is magnetically dominated throughout the cooling process \citep{tho04}.\\

The primary benefit of this model is that it can account for a late energy injection into the burst. This model predicts a plateau phase in its X-ray light curves, which is attributed to the spin-down of a newly formed magnetar \citep{zha06,troja2017a}. However, it poses a problem because the maximum amount of energy that can be extracted corresponds to the magnetar's rotational energy in this scenario. This can be calculated as \citep{lu14}
\begin{equation}
E_{rot}=\frac{1}{2}I\Omega_0^2=\frac{2 \pi^2 I}{P}\simeq 2.2\times10^{52}{\rm\ erg}\ M_{1.4}\ R_6^2\ P_{0,-3}^2\ ,
\label{eq:Erot_NS}
\end{equation}
assuming a canonical spherical NS with moment of inertia $I=\frac{2}{5}MR^2\simeq10^{45}\ \text{g cm}^{-2} \ M_{1.4}\ R_6^2$ with initial angular frequency $\Omega_0=\frac{2\pi}{P_0}$  and a period on a millisecond scale.\\

Even if some extreme massive values were taken into account, such as the hyper massive neutron stars near the \textit{Tolman-Oppenheimer-Volkoff} (TOV) limit, the maximum energy budget could not exceed $E_{\rm budget,max}=8.5\times10^{52}$ erg. This represents a problem because many GRB have been observed with energy reservoirs greater than this value. Typically, these GRB have been linked to a BH central engine capable of delivering such a large amount of energy.

\subsection{Fireball model}\label{subsec:fireball}
We rely on the most widely accepted model, known as the "fireball," to include GRB dynamics into our research. This model defines the interaction between the relativistic energy outflow and the GRB core engine. This notion demands the emission of a high concentration of radiation in a small amount of virtually baryon-free space. Because the temperature is higher than the rate of $e^\pm$ pair generation, nuclei are photodisintegrated, and the plasma is mostly made up of free $e^\pm$ pairs, $\gamma$-ray photons, and baryons. The base of the jet is generated by the so-called fireball plasma coupled to the progenitor. According to the fireball model, there will be two stages: the prompt emission: when jet inhomogeneities cause internal collisionless shocks \citep{1994ApJ...430L..93R,2017ApJ...848...15F} and the afterglow: when the relativistic outflow sweeps up enough external material. In terms of progenitor models, later light curve measurements point to a "compact" inner engine, which can be described using this fireball model independently of the progenitor emission mechanism considered.\\

Following the core collapse, the base of the fireball flow is linked to the GRB central engine. Initially, the fireball is opaque to neutrinos  $\left(\tau_{\nu_e}=54\ E_{52}^{5/4}r_{6.5}^{-11/4}\ \rm{and}\  \tau_{\nu_\mu}=7.4\ E_{52}^{5/4}r_{6.5}^{-11/4}\right)$\citep{koe05} but becomes transparent ($\tau_{\nu_e,\nu_\mu,\nu_\tau}<1$) as it expands and then neutrinos can escape.  Moreover, the fireball has strong magnetic fields and it is mainly composed of $e^\pm$ pairs and free nucleons that are basically at rest within the progenitor reference frame \citep{zha04}. A quasi-thermal equilibrium is reached ($\sim1 - 10$ MeV) within a typical size of $r=10^{6.5} - 10^7$ cm, and densities of $10^9\leq\rho\leq10^{12}$ g cm$^{-3}$\citep{pir99}.
During this phase, a large number of thermal neutrinos are created inside the fireball plasma due to the high temperature reached. Mainly, pair annihilation  processes dominate $\left(e^++e^-\to\nu_x+\bar{\nu}_x\right)$\footnote{The subindex $x$ denotes that neutrinos of any flavor can be produced during these reactions $x=e,\mu,\tau$  }. However, other reactions, such as nucleon-nucleon bremsstrahlung $\left(NN\to NN+\nu_x+\bar{\nu}_x\right)$, plasma decay $\left(\gamma\to\nu_x+\bar{\nu}_x\right)$, positron capture on neutrons $\left(e^++n\to p+\bar{\nu}_e\right)$, and electron capture on protons $\left(e^-+p\to n+\nu_e\right)$, are also crucial for effective cooling of the system \citep{dic72}. Since the latter reactions only produce neutrinos with a definite electron flavor, it is estimated that initially there is an over-proportion of this neutrino flavor. This effect will be considered later when we takes into account the initial flavor rates.
\\


\subsection{Magnetar central engine candidates}
Several authors have statistically acknowledged the differences between the two models in a broad sample of past GRB observations. The findings of the X-ray light curves, for the most part, favor the BH model, with only a few of them exhibiting the shallow plateau. Over time, a collection of potential GRB candidates that satisfy the magnetar model has been compiled. For instance, \citep{ber13} presented a compendium of eight GRB with known redshift discovered by \textit{Swift/BAT} between 2005 and 2009 that, in addition to showing signs of late activity in the form of a plateau, also showed signs of precursor activity. This condition can only be explained through accretion processes if the central engine corresponds to a newly formed magnetar.\\

\citep{lu14} showed a sample of 214 magnetar central engine candidates with known redshifts and classified them into four groups (Gold, Silver, Aluminum, and others) based on the probability that a magnetar could produce one of these bursts. The Gold sample consists of three GRB with a well-defined internal plateau followed by a steep decay with a slope greater than three in their light curve. The Silver sample is made up of GRB that exhibit a shallow decay followed by a normal decay phase and also satisfy the $\alpha_1-\alpha_2$ "closure relation" of the standard external shock model \citep{gao13,wan15}, where $\alpha_1$ represents the time index during the plateau phase, and $\alpha_2$ corresponds to the time index of either the subsequent normal decay or the steep decay.  They also include a series of magnetar central engine candidates that could power a sGRB, as well as those that have a low probability of being magnetars but do not meet one or more of the conditions mentioned above. \\

On the other hand, \citep{li18} used a similar logic to subclassify candidates from the \textit{Swift/XRT} light curves. Their sample consists of 101 progenitors with an external plateau divided into Gold, Silver, and Bronze and where the criteria shown in Table \ref{tab:li18_classification} was followed.   Here $E_{\rm X,iso}$ denotes isotropic X-ray energy, $E_{\rm K,iso}$ represents isotropic kinetic energy, and $E_{\rm budget}$ connotes the maximum budget energy released by a magnetar. As Equation \ref{eq:Erot_NS} reveals, this value is $E_{\rm budget}\sim2\times10^{52}$ erg. According to the authors, roughly 23\% of the candidates (Bronze) show signs of having a magnetar as a central engine, whereas the remaining belong to BH progenitors (Gold and Silver).\\

\begin{table}[htbp!]
\begin{tabular}{@{}ccc@{}}
\toprule
\textbf{Type}   & \textbf{Condition}                                  & \textbf{Progenitor} \\ \midrule
\textbf{Gold}   & $\displaystyle\{E_{\rm X,iso},E_{\rm K,iso}\displaystyle\} >E_{\rm budget}$ & BH  \\           
\textbf{Silver} & $E_{\rm X,iso}<E_{\rm budget}<E_{\rm K,iso} $ & BH \\
\textbf{Bronze} & $\displaystyle\{E_{\rm X,iso},E_{\rm K,iso}\displaystyle\} <E_{\rm budget}$ & Magnetar \\
\bottomrule
\end{tabular}
\centering
\caption{Summary of the criteria used by \citep{li18} to classify GRB progenitors. }
\label{tab:li18_classification}
\end{table}

\section{Results}\label{sec:results}
\citep{fra16} showed that the resonance lengths for MeV-neutrinos were between $1.3\times10^{6}$ and $7.6\times10^{7}$ cm for neutrinos with energy ranging from 5 to 30 MeV, considering the three-flavor admixture scenario within a plasma fireball. In this work, we demonstrate that the action of this medium will have a significant effect on the neutrino oscillation dynamics during the development of a LGRB.\\

With this in mind, we investigate how the neutrino potential associated with a $\sim 10^7$ cm-sized fireball varies under several conditions. We decided to divide the study into two cases for this purpose. The first one will be called "BH" an this will refer to a black hole precursor with a typical magnetic field of $B=10^{12}$ G. Similarly, "magnetar" shall refer to those LGRBs formed by a magnetar with $B\sim10^{14}-10^{15}$ G, which is a good approximation that fits the current observations of magnetar's field strengths \citep{metzger10,kas17}.\\

\begin{figure*}[htbp!] 
\centering
	\subfloat (a)
					{
  				\includegraphics[width=0.45\textwidth]{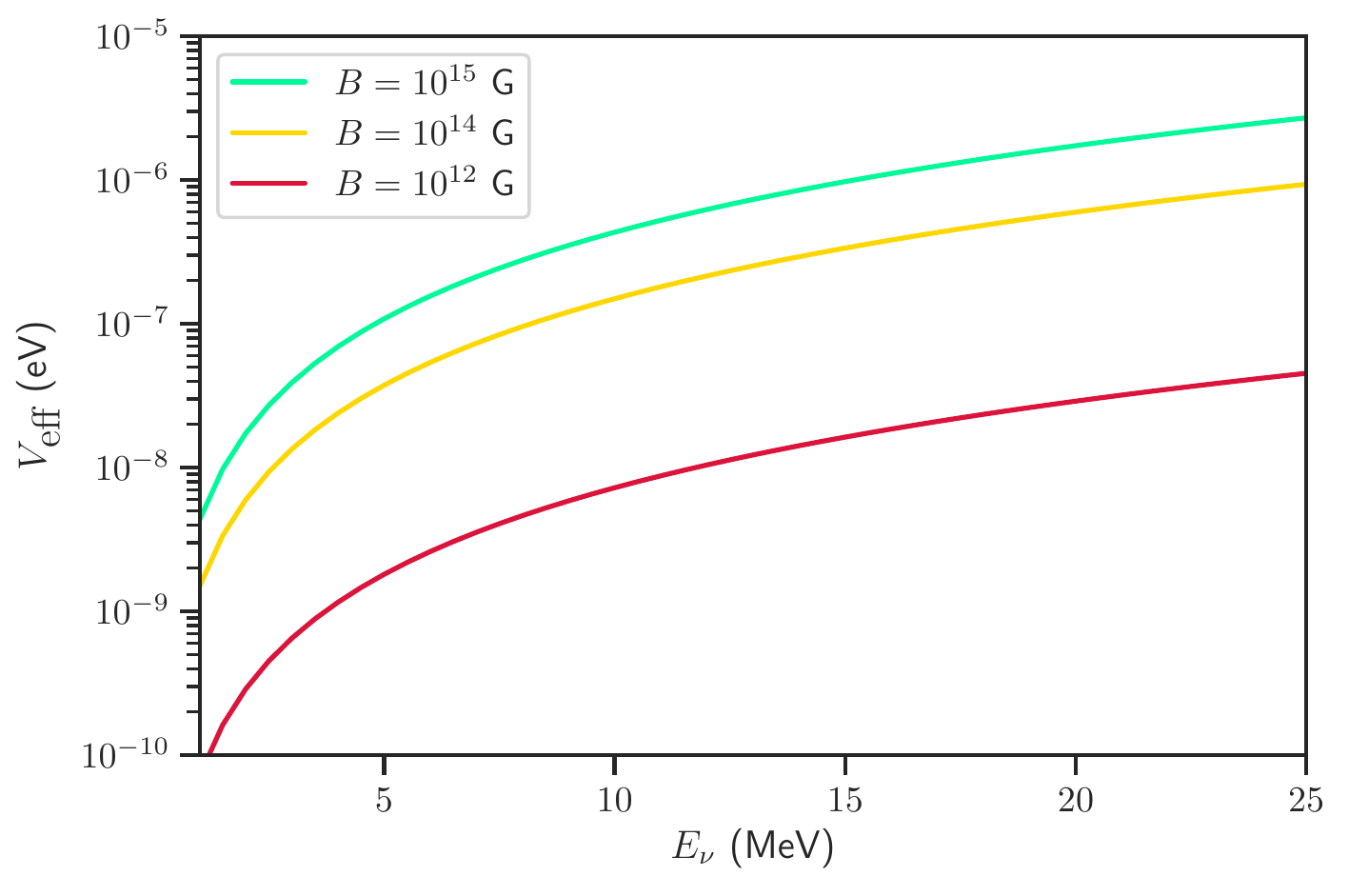}%
 				    \label{plot:V_Enu}
          	        }
	\subfloat (b) 
					{
  				\includegraphics[width=0.45\textwidth]{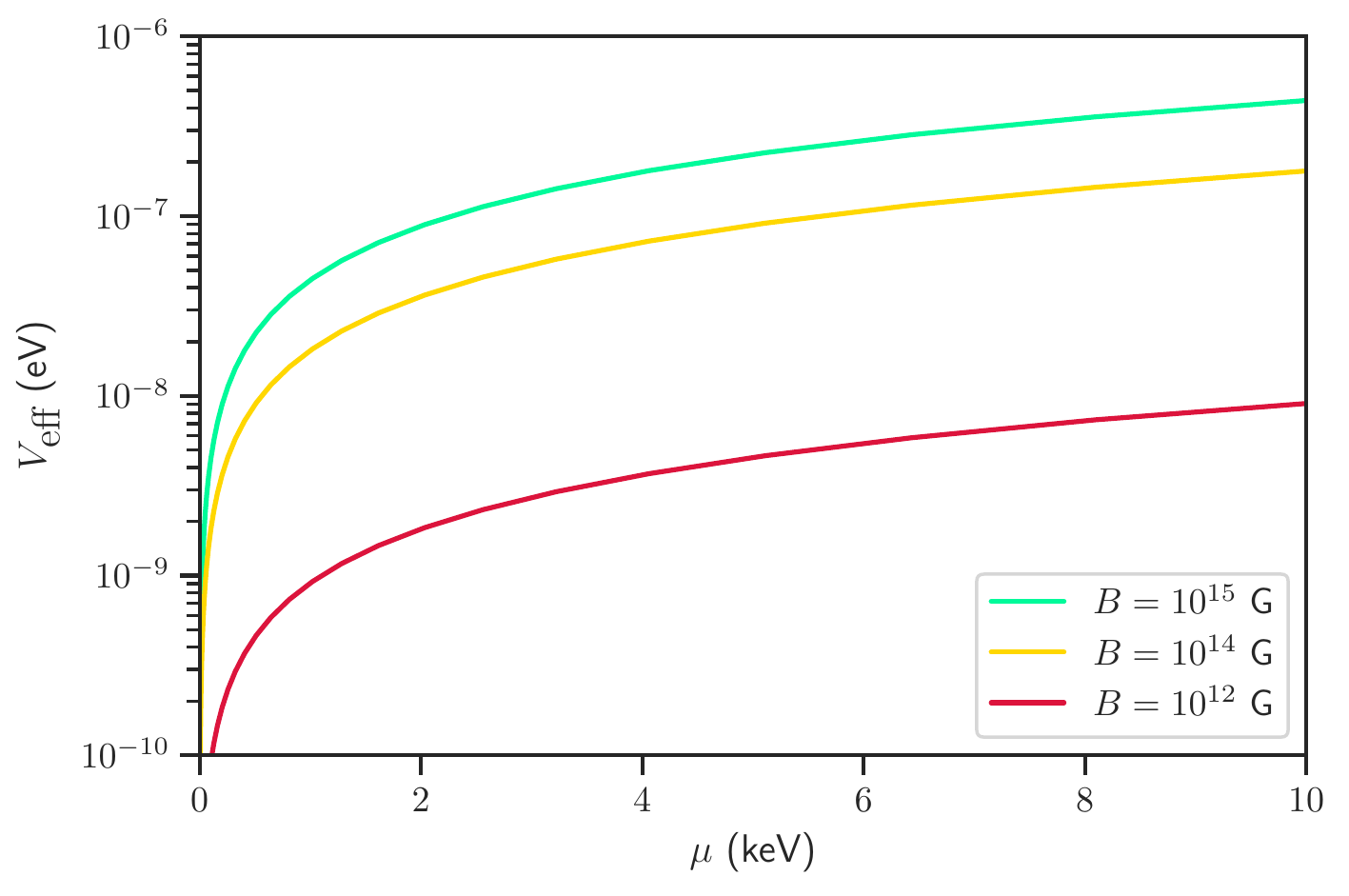}%
 				    \label{plot:V_mu}
          	        }    
          	        \qquad  	       
	\subfloat (c)
					{  					\includegraphics[width=0.45\textwidth]{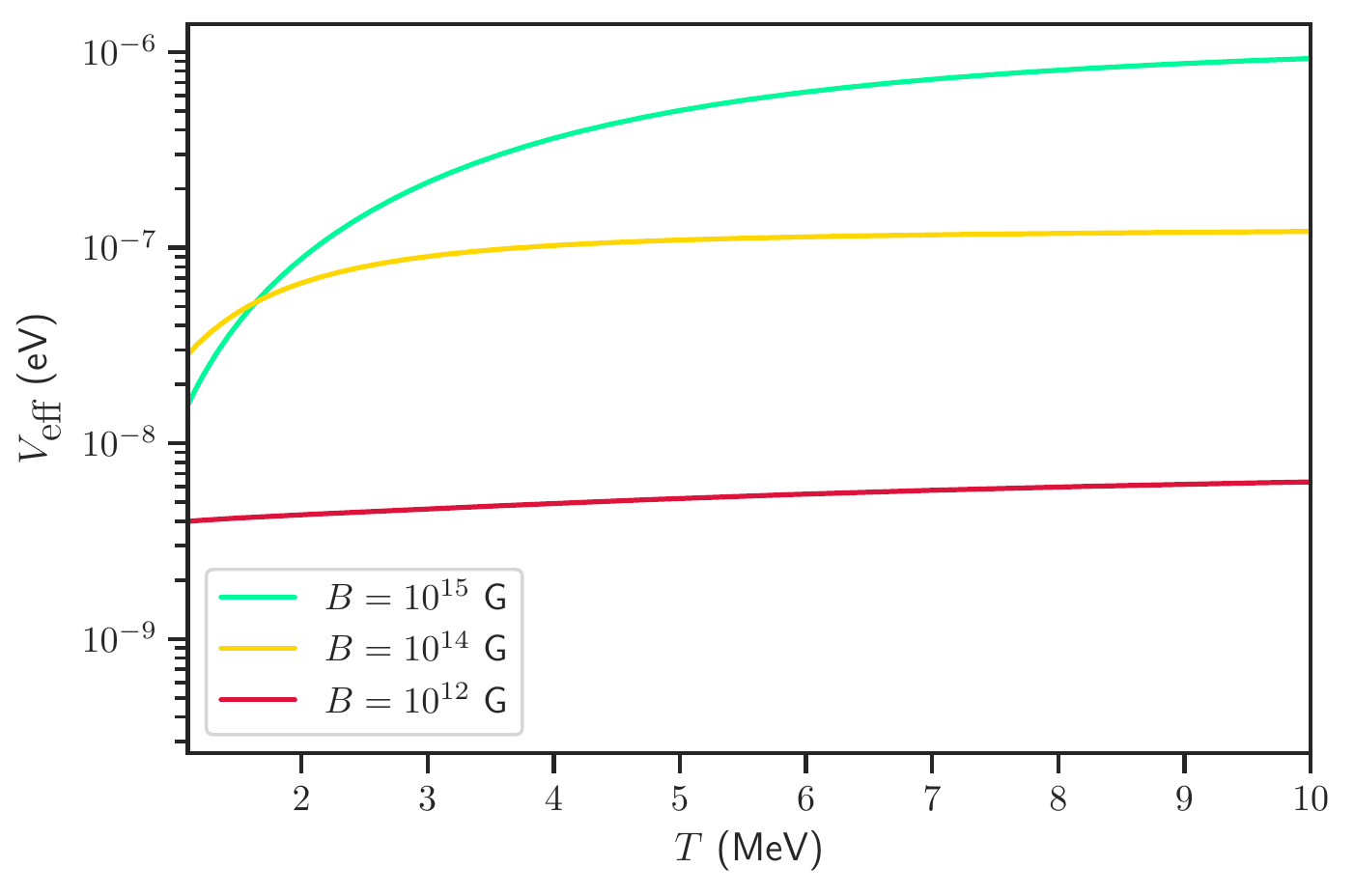}%
 				    \label{plot:V_T}
          	        }
	\subfloat (d)
					{
  				\includegraphics[width=0.45\textwidth]{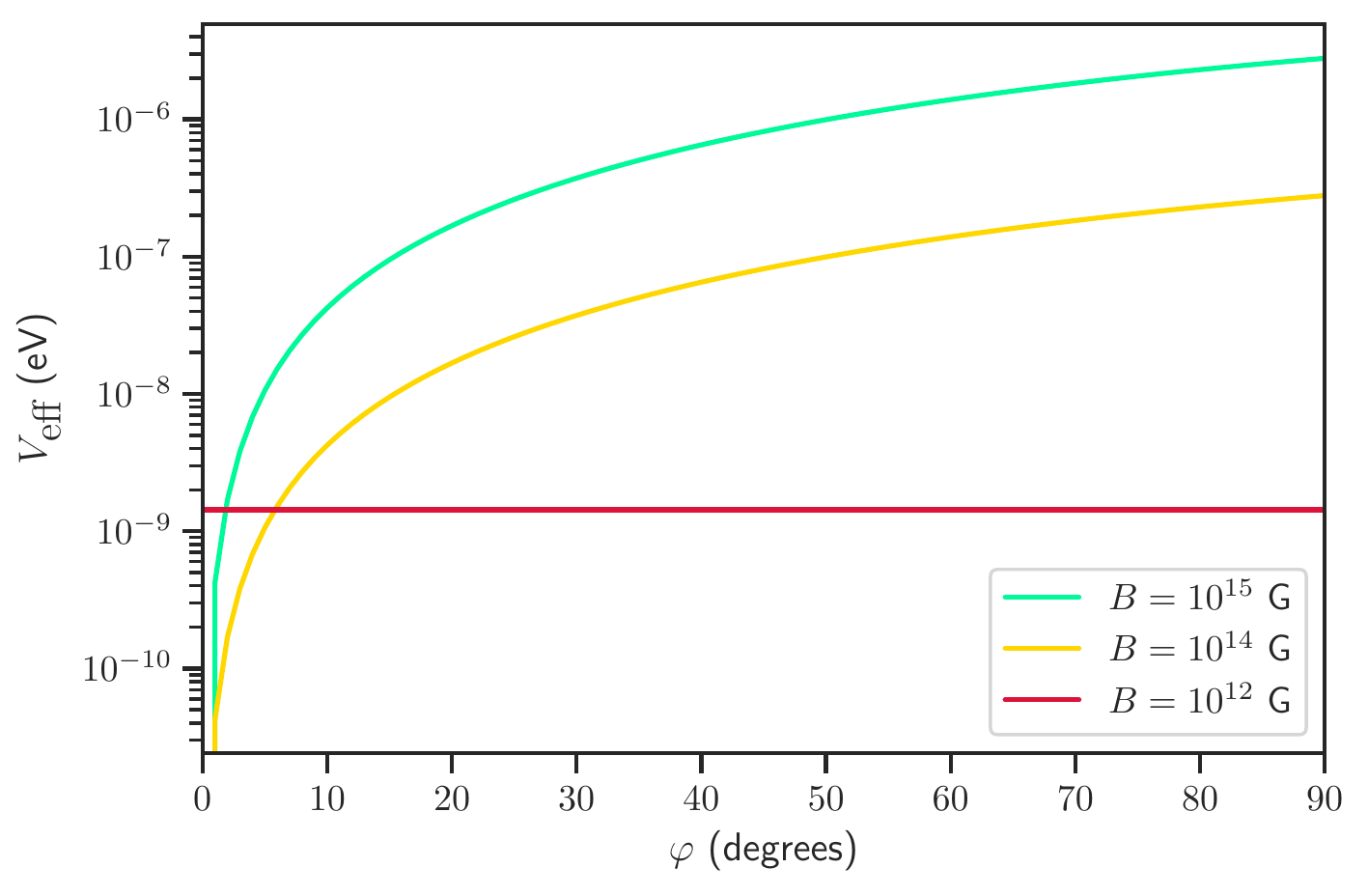}%
 				    \label{plot:V_phi}
          	        }   
\caption{Representation of neutrino potential contributions for magnetic field values of $10^{15}\ \rm G$ (green line), $10^{14}\ \rm G$ (yellow line), and $10^{12}\ \rm G$  (red line). Since the effective potential is a multivariable function, we plot it as a function of neutrino energy (top left), chemical potential (top right), temperature (bottom left), and neutrino propagation angle (bottom right). {When we represent the neutrino effective potential  in terms of a single variable, we set the other parameters to their typical values for a fireball . These values are: $E_\nu = 20$ MeV, $\mu = 5$ keV, $T = 3$ MeV, and $\varphi =  0\ ^\circ$.}  In these figures, we can see that the potential is an increasing function that reaches higher values for stronger magnetic fields, making this variable one of the main causes that perturb and increase the value of the potential inside the fireball. }
\label{fig:Veff}
\end{figure*}

Our goal is to examine the changes in neutrino oscillation attributes that occur as they traverse through a fireball with different magnetic fields. In the magnetar scenario, we computed the potential within a strong $\vec{B}$ limit $(\Omega_B>1)$ using  Equation \ref{eq:veffs}, while in the BH case, we calculated the potential within a mild $\vec{B}$ limit $(\Omega_B<1)$ using  Equation \ref{eq:veffw}.  Because both equations are multivariable, we exhibit their dependency on neutrino energy, temperature, chemical potential, and propagation angle in Figure \ref{fig:Veff}.  We adopt the typical values associated with a $10^7$ cm--sized fireball with the following characteristics: ($1\rm\ MeV\leq E_\nu\leq 30\rm\ MeV$; $5\times10^{-2}\rm\ keV\leq \mu\leq 10\rm\ keV $; $1 \rm\ MeV\leq T\leq 10 \rm\ MeV $; and $0^\circ\leq\varphi\leq 90^\circ $) in both scenarios \citep{goo86,pac93,pir99,bel03,mes06}. Figure 1 shows that the potential is an increasing function that is highly dependent on magnetic field variations and lies in the range between $3\times10^{-9}$ eV and $7\times10^{-7}$ eV.\\

\begin{figure}[htbp!]
    \centering
    \centering
	\subfloat (a)
					{
  				\includegraphics[width=0.45\textwidth]{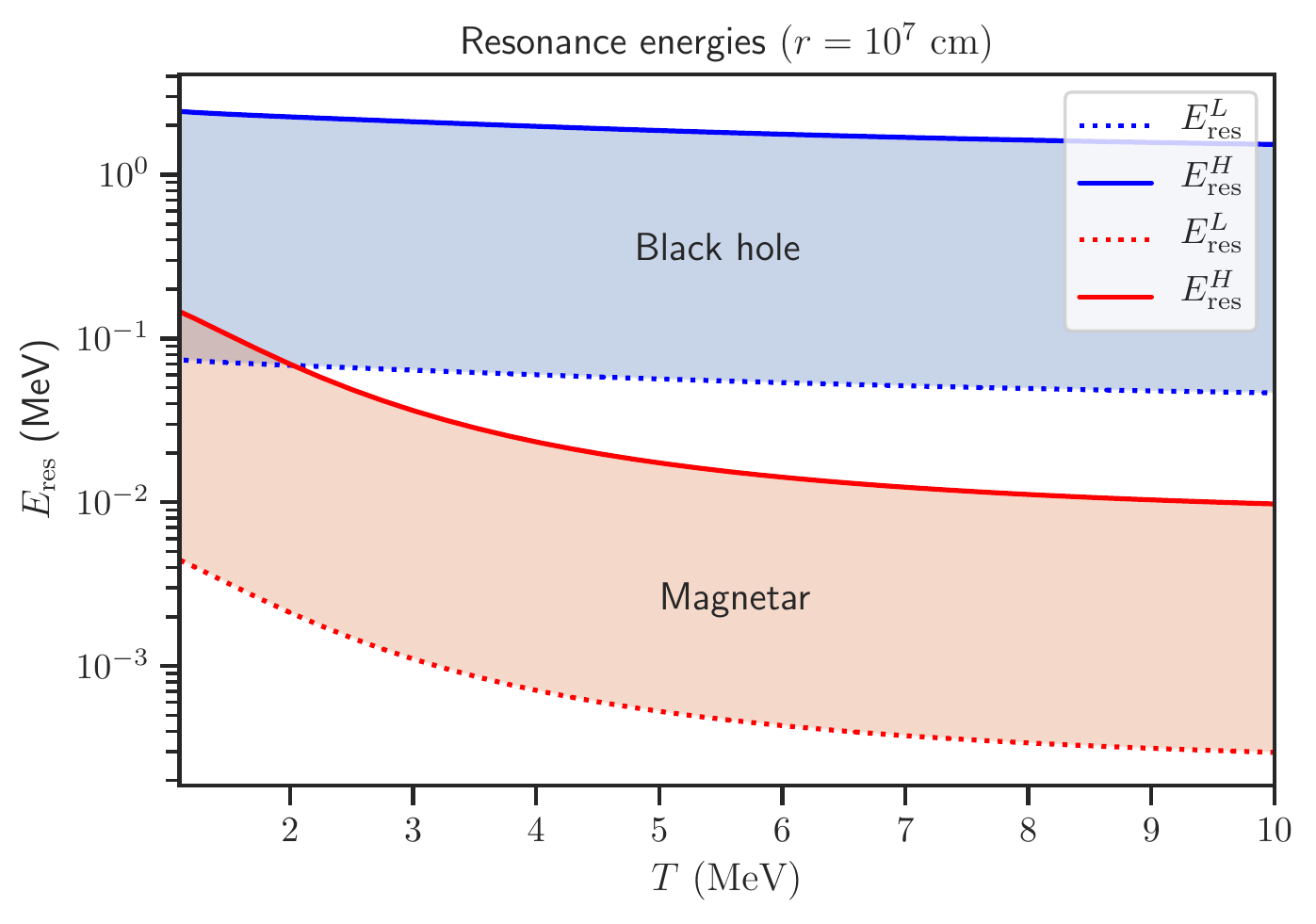}%
 				    \label{plot:Res1}
          	        }
	\subfloat (b) 
					{
  				\includegraphics[width=0.45\textwidth]{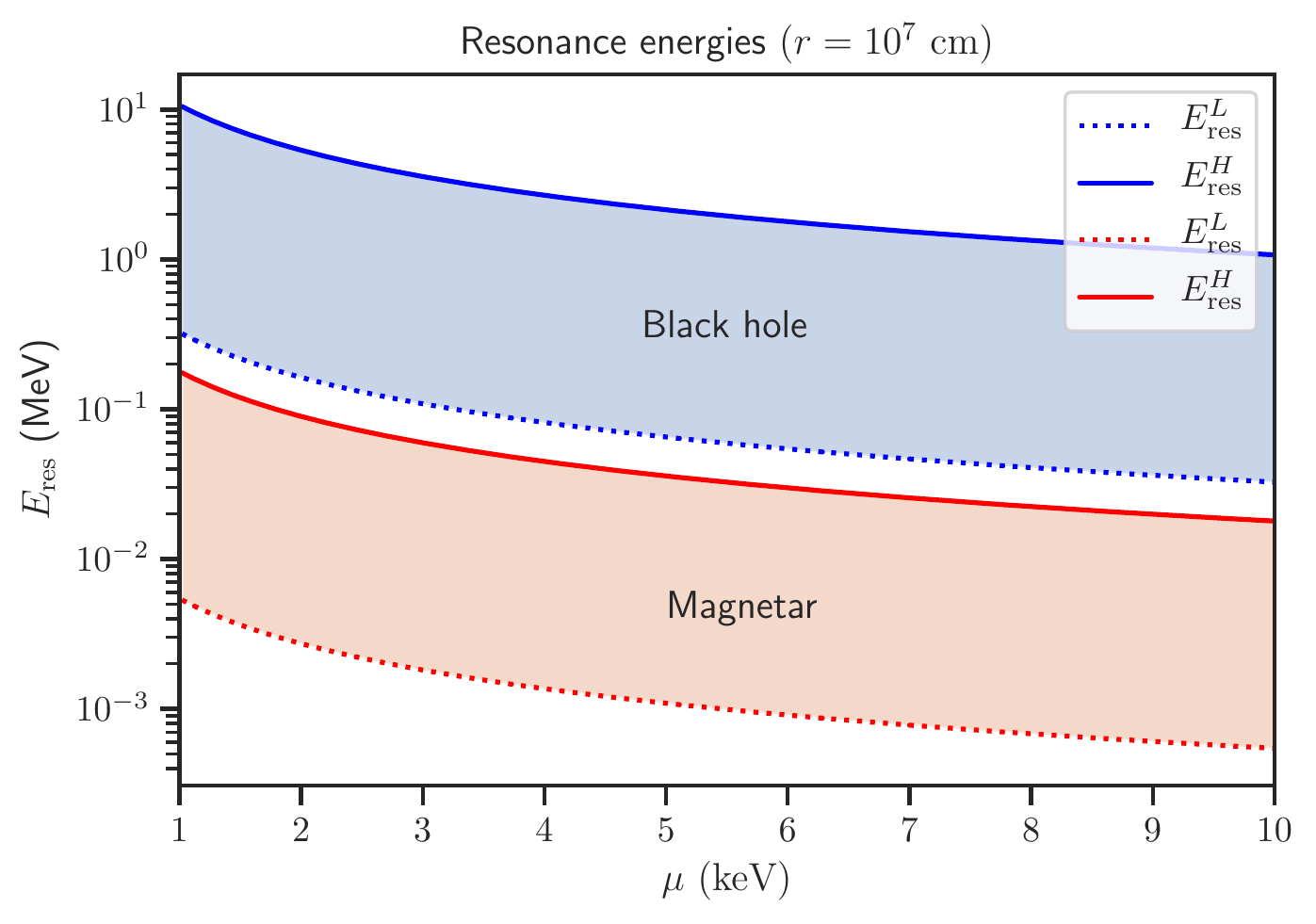}%
 				    \label{plot:Res2}
          	        }    
          	        \qquad

\caption{\\
\textbf{Left:} Temperature-dependent resonance energy ranges for the two types of progenitors considered: a BH (blue region) and a magnetar (red region). The lower limits are represented by a dotted line, while the upper limits are displayed by a solid line. In order to make this plot, we have set the other values to  $\mu=5$ keV, $\varphi=0\ ^\circ$ and $E_\nu=20$ MeV.\\
\textbf{Right:} The same as the left-hand figure, but as a function of chemical potential.  The medium temperature is set to 3 MeV.\\
We can see that the condition $E_\nu>E_{\rm res}^L$ is fulfilled in both cases for thermal neutrinos with typical energies lying in the MeV-range, implying that resonance effects within this medium will be significant.}
\label{fig:Res_energies}
\end{figure}

We calculate the resonance energies for both scenarios using Equation \ref{eq:res_energy} to justify that neutrinos will oscillate resonantly throughout the entire range of energy covered. As a result, we display these resonance energies in Figure \ref{fig:Res_energies} while taking into account the effective potential associated with a BH (blue region) and a magnetar (red region). In both cases, the relation $E_\nu>E_{\rm res}^L$ is fulfilled for the whole MeV spectra and therefore the effects of the material medium become prominent.\\

\begin{figure}[htbp!]
    \centering
    \includegraphics[width=0.88\textwidth]{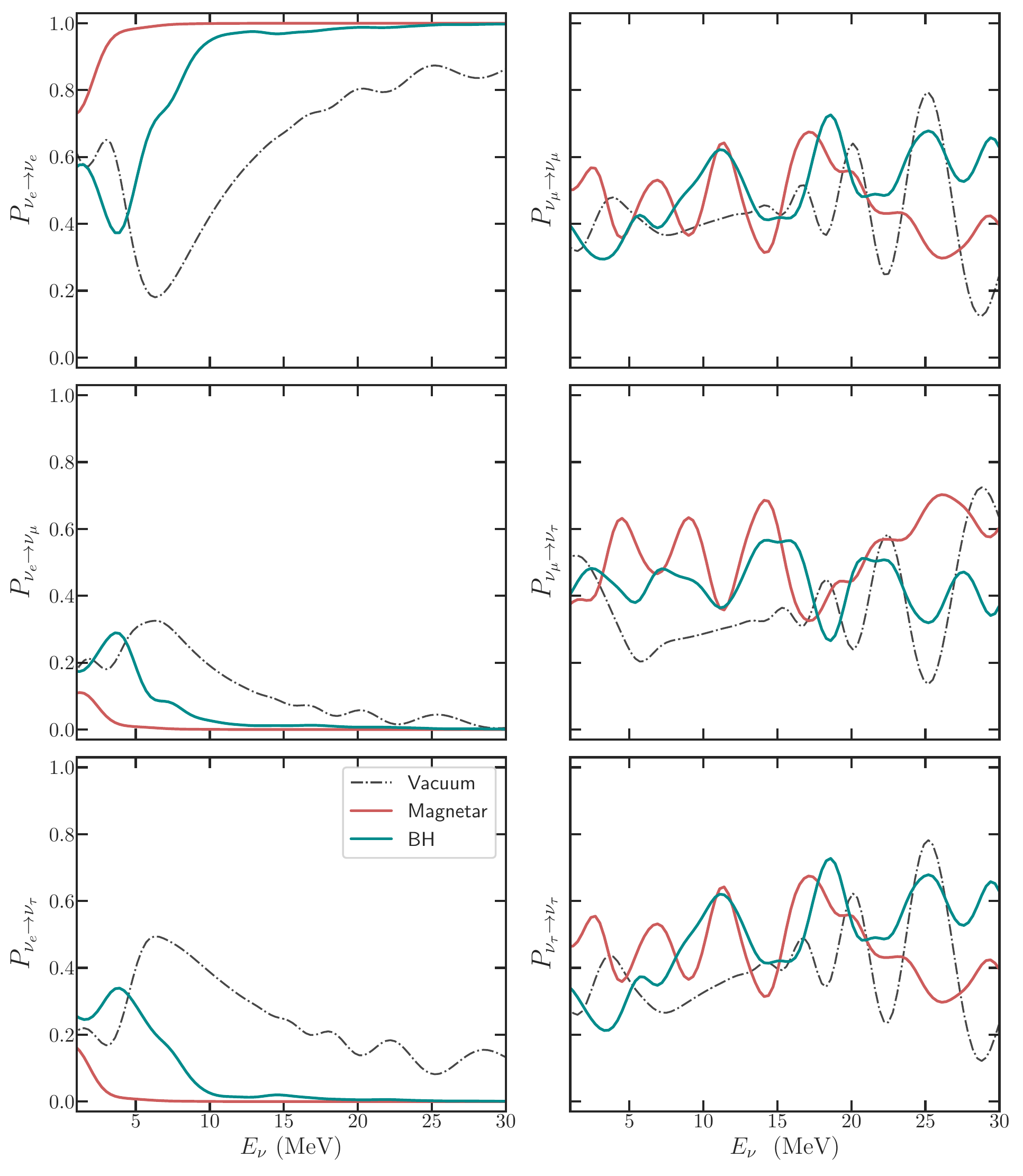}
    \caption{ Oscillation probabilities as a function of energy for neutrinos traversing a fireball through a magnetic field attributed to a BH (dark cyan line), a magnetar (red line), and a comparison with theoretical probabilities expected in the vacuum (gray dashed line). Only the six independent transitions permitted within the NO scheme are depicted.
}
    \label{fig:Probas}
\end{figure}

Once we have identified the potential for both specific cases, we can build the related hamiltonian that incorporates all  the  characteristics of the medium, allowing us to establish the neutrino state of Equation \ref{eq:Sch_solution}.  Then the probability of oscillation between two neutrino states with defined flavor is determined by solving numerically Equation \ref{eq:Proba_m}. Figure \ref{fig:Probas} depicts these results as a function of energy when the oscillation parameters of three flavors inside a NO scheme are considered.  It is worth mentioning that only the six independent transitions are shown since $P(\nu_\mu\to\nu_e)=P(\nu_e\to\nu_\mu),\ P(\nu_\tau\to\nu_\mu)=P(\nu_\mu\to\nu_\tau), \ P(\nu_\tau\to\nu_e)=P(\nu_e\to\nu_\tau)$. for simmetry.  The magnetar scenario is shown in red, while the BH case is highlighted in blue. We also compare the theoretical oscillation probabilities computed with the Hamiltonian in the vacuum, which are indicated in gray dotted lines.\\

From Figure \ref{fig:Probas}, we observe  that because the magnetic field has a significant influence on the potential, the oscillation probabilities deviate from the expected value in the vacuum. This effect is more pronounced in the magnetar case (red line), where the magnetic field is stronger, and for all $\nu_e$-flavor transitions (figures in the left column), where the oscillation phases shift.
The probabilities in the right column show growing wiggles at low energies in the same region where a plateau occurs in the vacuum situation. \\

\begin{figure*}[htbp!] 
\centering
	\subfloat
					{  					\includegraphics[width=0.45\textwidth]{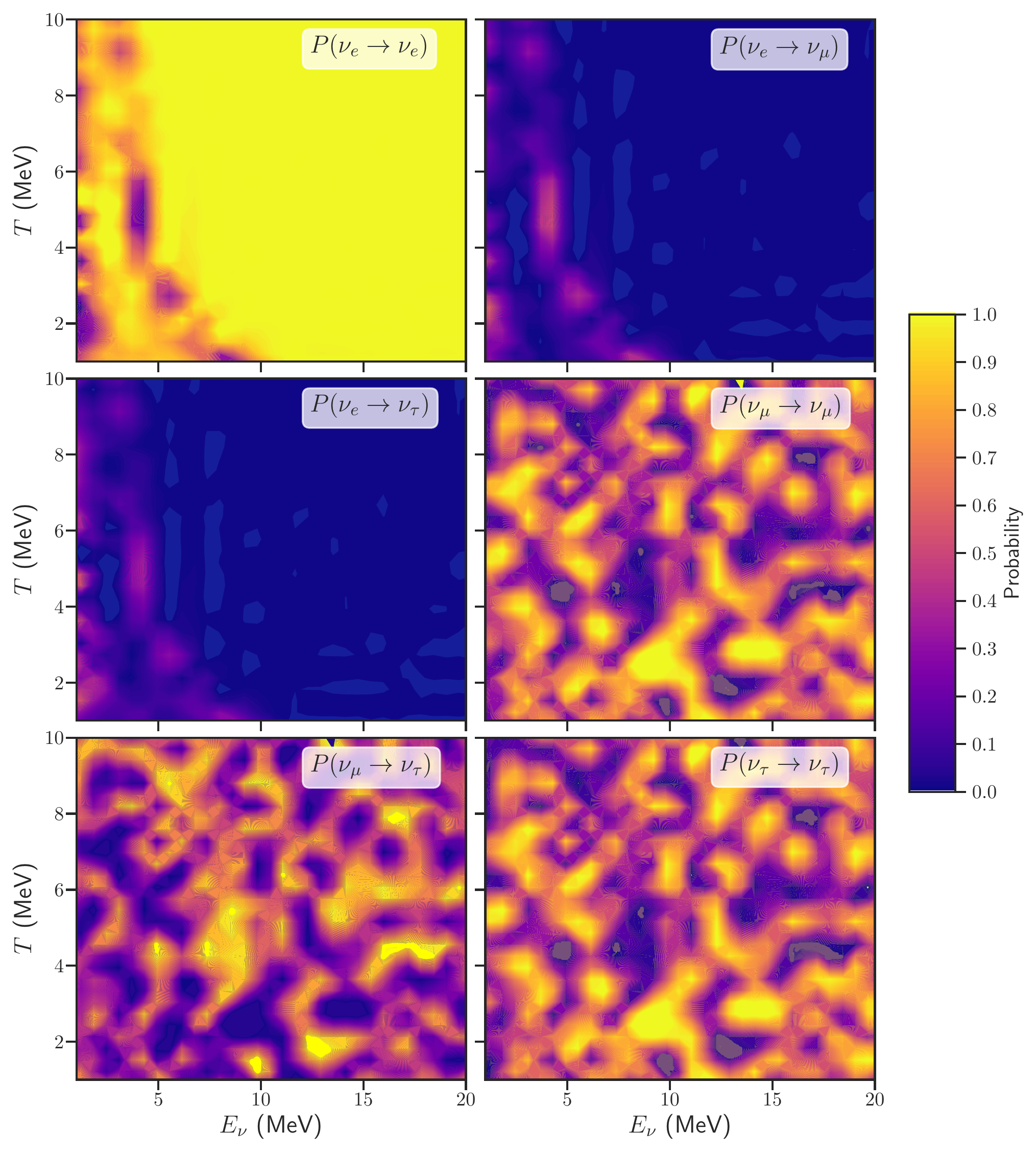}
 				    \label{plot:P_BH}
          	        }
					\qquad
	\subfloat
					{
  					\includegraphics[width=0.45\textwidth]{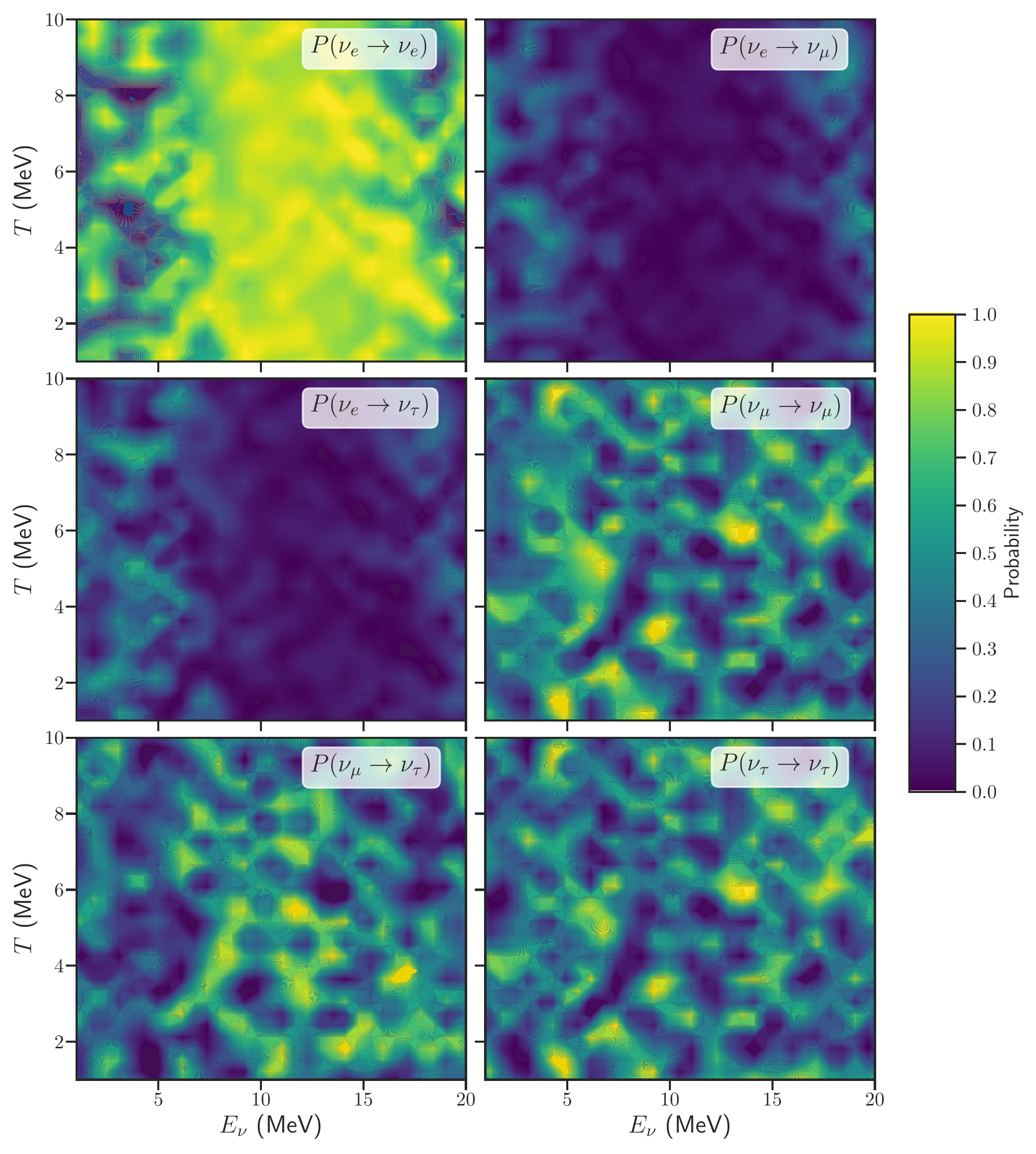}%
 				    \label{plot:P_BNS}
          	        }    
          	        \qquad  
\caption{(Left) Neutrino oscillograms constructed on the ($T-E_\nu$) plane for neutrinos propagating within a magnetic field of $B=10^{12}$ G  (BH), and a magnetar $B=10^{15}$ G (right). In both cases, In both cases, we notice a prevailing survival of the electron neutrino.
}
\label{fig:Oscillograms}
\end{figure*}

In order to account for other possible dependencies in the calculation of these probabilities, we have included the relevant oscillograms for the magnetar (left side) and the BH (right side) in Figure \ref{fig:Oscillograms}, where we incorporate contour plots for the energy and temperature variables. We notice that the survival of the electron neutrino predominates in both scenarios, which can be attributed to the fact that this is the only flavor that interacts with the media's leptons via CC and NC, whilst the other two flavors exclusively interact via NC.\\

\begin{figure*}[htbp!]
\centering
	\subfloat
					{  					\includegraphics[width=0.45\textwidth]{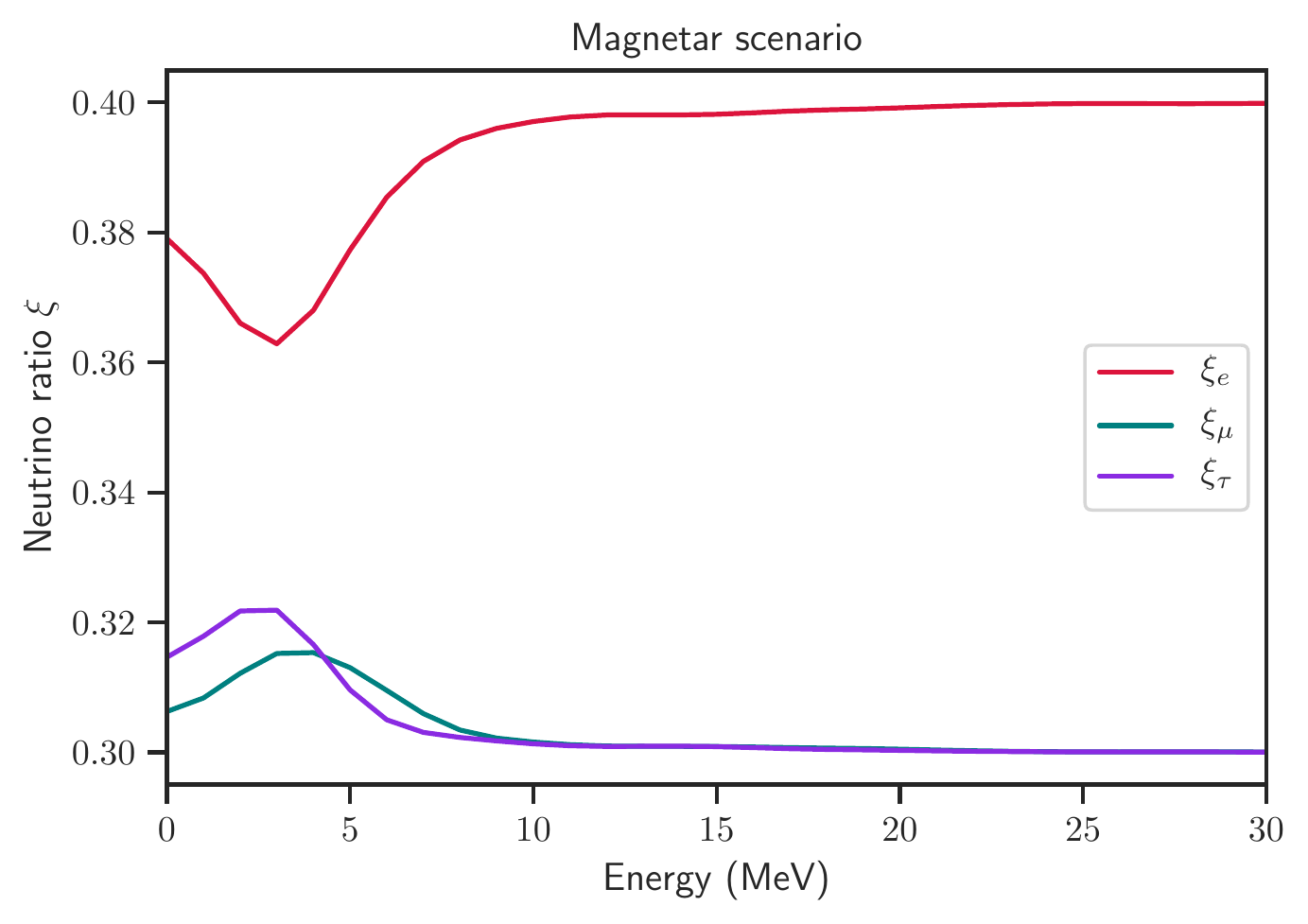}
 				    \label{plot:ratio_BH}
          	        }
					\qquad
	\subfloat
					{
  					\includegraphics[width=0.45\textwidth]{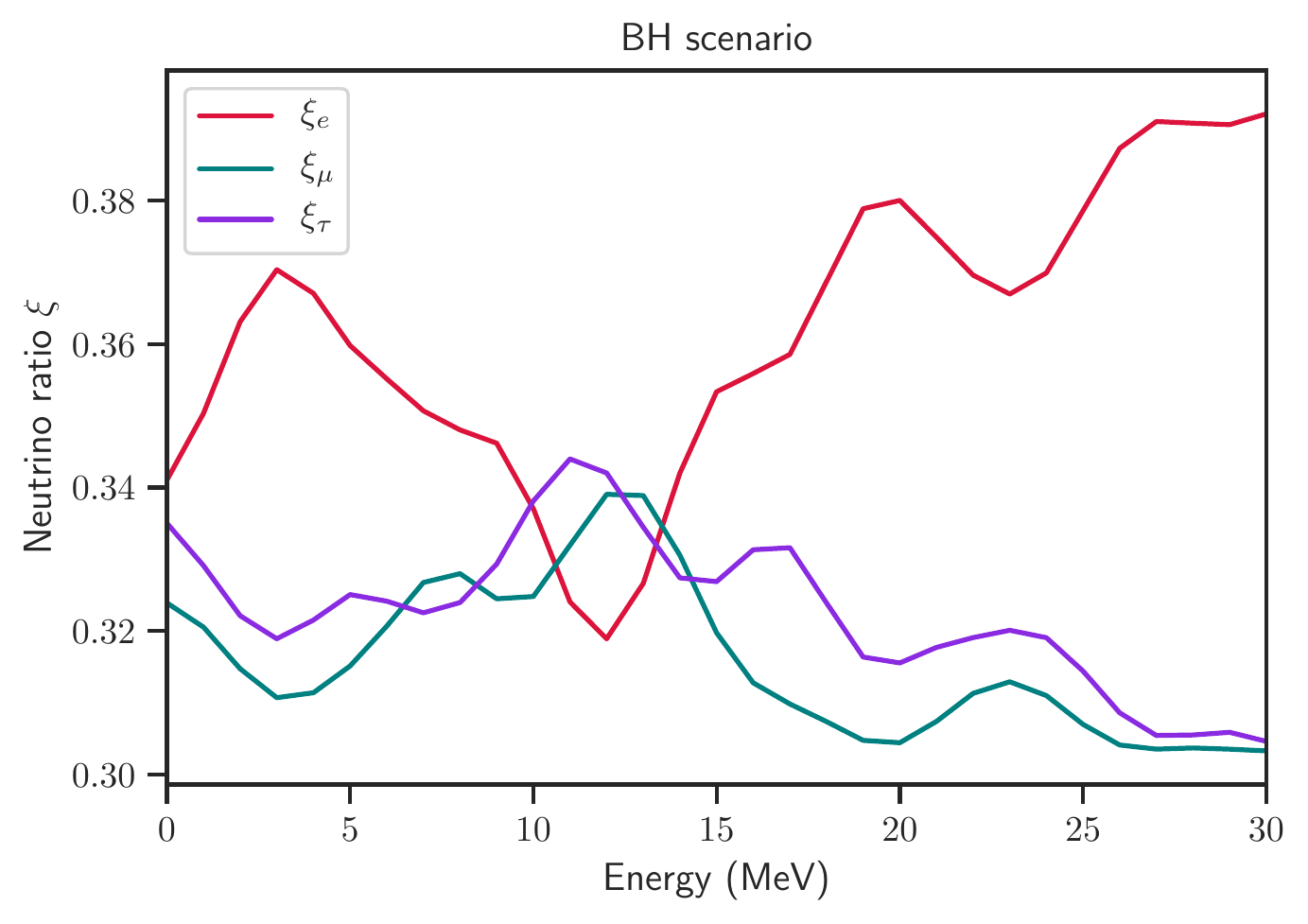}%
 				    \label{plot:ratio_N}
          	        }    
          	        \qquad  
          	        
	\subfloat
					{
  					\includegraphics[width=0.45\textwidth]{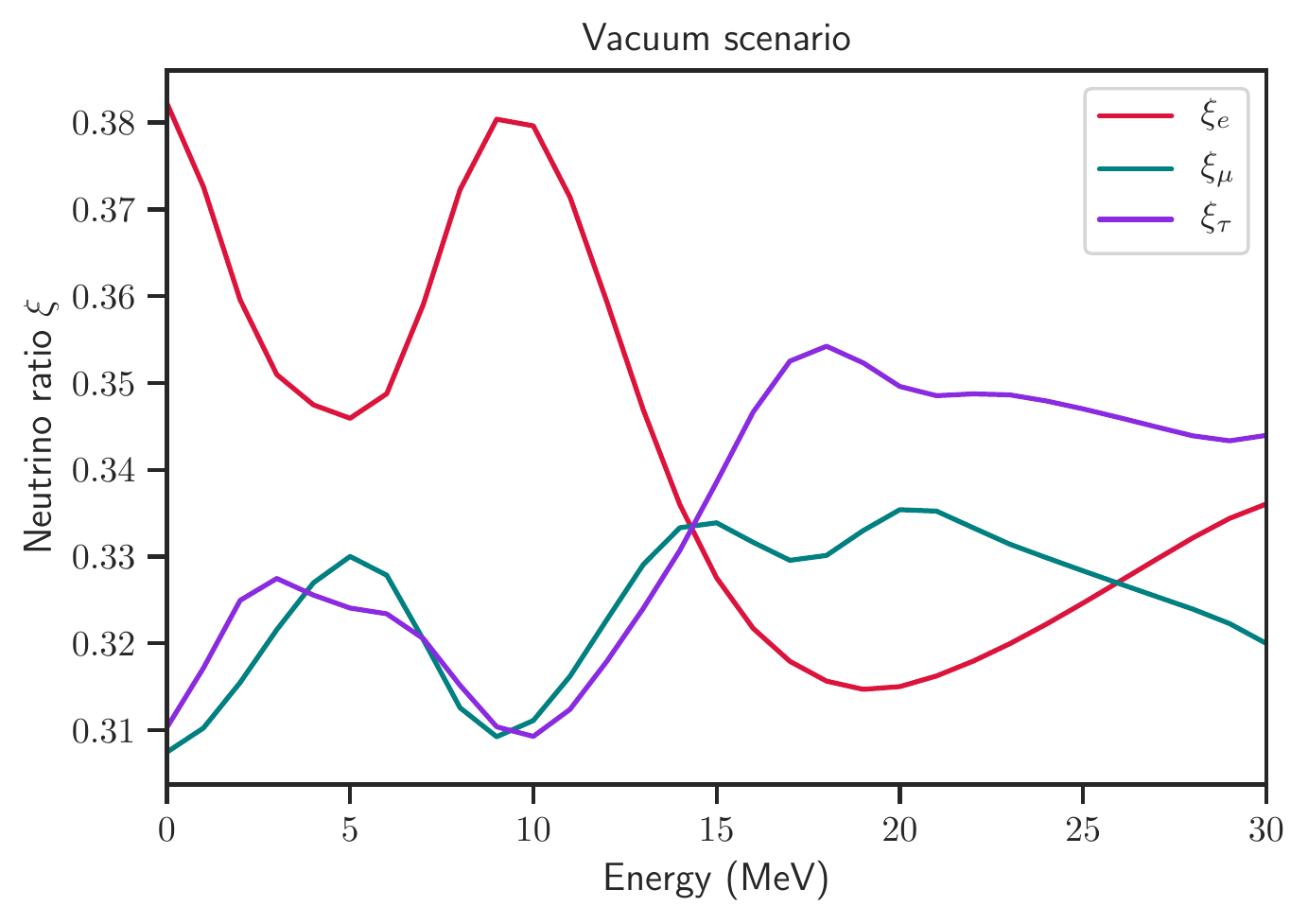}%
 				    \label{plot:ratio_V}
          	        }    
          	        \qquad  

\caption{Individual flavor ratios expected in ground-based detectors as a function of neutrino energy for {the magnetar scenario (top-left), BH scenario (top-right)}, and the vacuum case for comparison (bottom). The flavor corresponding to the electron neutrino is shown in magenta, the flavor corresponding to the muon neutrino in purple, and the flavor corresponding to the tau neutrino in blue.}
\label{fig:ratios}
\end{figure*}

As previously stated in Section \ref{subsec:fireball}, several thermal processes are responsible for producing these multi-MeV neutrinos, where the capture of $e^\pm$ in the nucleus is the only reaction capable of producing neutrinos with a defined electron flavor. Therefore, to account for this effect in our calculations, we chose an initial creation rate of $\xi_{\rm created}\equiv (\nu_e:\nu_\mu:\nu_\tau)=(0.4:0.3:0.3)$, assuming a slight initial overproduction of $\nu_e$. With this in mind, we can now use Equation \ref{eq:xi_n} to calculate the fraction of neutrino flavors that leave the source as a function of $E_\nu$ (Eq. \ref{eq:xi_n}). These results are shown together in Fig. \ref{fig:ratios}. When we contrast them to the theoretical fractions expected in the vacuum, we notice that the two outputs are not identical. For instance, the predicted neutrino rate in the magnetar scenario approaches the proposed starting rate at high energies, but it is more fluctuating in the BH's case. As a result, combining these findings would allow us to determine, for example, the initial magnetic field conditions that the remnant faced during neutrino emission. This would serve as an additional mechanism to determine the central engine model when other discrimination mechanisms are inconclusive or to confirm them within the neutrino channel.  \\

\begin{figure}
    \centering
    \includegraphics[width=0.95\textwidth]{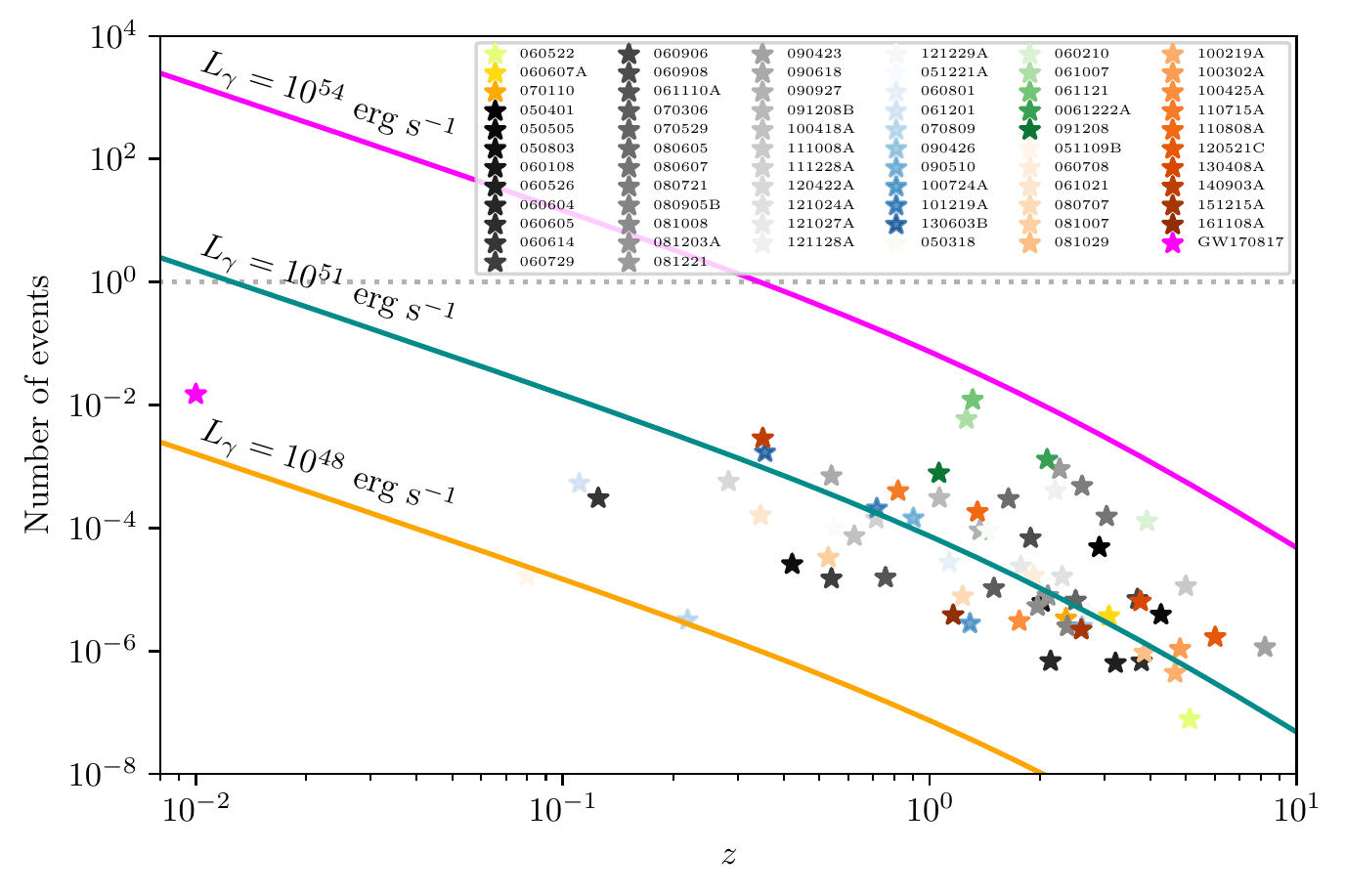}
    \caption{ The number of 20-MeV neutrinos expected in the Hyper-K detector  from GRB events with known redshift.  Colored lines represent the isotropic-equivalent photon luminosity of such events.
    The GRB sample is organized in a color code described below. (The first two classifications were made by \citep{lu14})\\
    i) Yellow sequential color scale or \textit{Gold}: LGRB with an internal plateau and a steeper decay.\\
    ii) Gray scale or \textit{Silver}: similar to the previous case, it represents those bursts with a shallow decay but with the closure relation ($\alpha_1-\alpha_2$) preserved.\\ 
    iii) Orange scale or \textit{Bronze}: LGRB whose emitted isotropic energy is less than the magnetar's budget energy ($E_{\rm budget} <2\times10^{52}$ erg) \citep{li18}.\\
    iv) Green scale: classification of some LGRB with the same properties as previously described made by \citep{ber13}.\\
    v) Blue scale: Sample of short GRB with known redshift \citep{lu14}.\\
    vi) Magenta: The GRB/GW 170817 event.\\}
\label{fig:Nevents_Magnetar}
\end{figure}

Assuming in Equation \ref{eq:Lg_Lnu} that the neutrino and electromagnetic fluxes are proportional ($L_\nu\sim10^2L_\gamma$), we estimate by Equation \ref{eq:Nev}, the number of neutrinos expected to be observed with the Hyper-K detector from these sources. So Figure \ref{fig:Nevents_Magnetar} displays these results from LGRB with known redshifts and whose progenitor has also been linked to a magnetar based on their X-ray light curves \citep{ber13,lu14,li18}. As a function of redshift, this figure illustrates the number of expected incident neutrinos and the isotropic luminosities ($L_{\gamma, \rm iso}$) associated with each event. Also shows that the sample belongs to extragalactic sources {with a redshift greater than 0.1}, which, despite being extremely energetic, were too far away to be detected by this experiment.\\

However, we might argue that the combination of a sufficiently luminous source located close to the Earth may be conducive to the detection of  such particles. For instance, in the past, events of interest such as GRB 170817A, (where the gravitational wave plus electromagnetic counterpart of a short GRB was detected for the first time and whose associated progenitor was attributed to the merger of two neutron stars), were located in spatial correlation within the host galaxy NGC 4993 at a distance of $\approx 40$ Mpc \citep{abott17}, which is equivalent to a $z\approx10^{-2}$. {We estimate that we could detect about 20 neutrinos with the Hyper-K detector at these distances, but for sources with a typical luminosity of $L_\gamma\gtrsim10^{51}$ erg s$^{-1}$.
Figure \ref{fig:Nevents_Magnetar} also suggests that the minimum distance required to detect a single neutrino for a very bright source ($L_\gamma=10^{54}$ erg s$^{-1}$) is $z\sim0.3$, while for a redshift of $z\sim0.01$, this number rises dramatically to $\sim 4\times10^3$ neutrinos.
However, because the latter is the best-case scenario, it is less likely to occur.}\\

Since the dynamics of short and long GRB are similar in terms of thermal neutrino production. We can be optimistic that an LGRB with a typical luminosity of  $(10^{51}-10^{52}\ \rm erg\ s^{-1})$, spotted at a sufficiently close distance as the GW170817 event, will significantly raise the likelihood of detecting neutrinos. \\

\section{Conclusion}\label{sec:conclusion}
Neutrinos by nature are quite remarkable particles that could explore the astrophysical sources (e.g. a magnetar or a BH) that release them,  owing to the fact that these originate in a very hot, magnetized, and initially opaque medium. They  travel towards the Earth through a dense column density. In this paper, we showed that once neutrinos transit a non-vacuum medium, their oscillation probabilities are affected by the surrounding medium but mainly by the magnetic component. Therefore, we found a mechanism to identify the central engine associated with these events by studying and recognizing these variations.\\

We illustrate the significance of the magnetar model's  magnetic contribution within a fireball with a $\vec{B}$ endowed to a magnetar via the effective neutrino potential and contrast them with the BH case, where this effect is not as prevalent. We found that even when the potential is reliant on numerous physical variables,  the magnetic field contributes the most to the potential and constitutes more than one order of magnitude larger in the magnetar scenario. Furthermore, we identified that these neutrinos will leave the fireball with a predominant $\nu_e$ survival rate. \\

The probabilities provide a preliminary approximation of how these particles behave in each media type. Whereas the expected neutrino rate, in this context, gives a greater understanding of this impact in observable terms. We found, in particular, that the probabilities differ somewhat from the predicted likelihood in the vacuum. For illustrate, the survival probability for a 12 MeV electron neutrino traveling within a fireball in: i) a magnetar is $P(\nu_e\to\nu_e)=0.999934$, ii) BH case is $P(\nu_e\to\nu_e)=0.937288$, and iii) the vacuum is $P(\nu_e\to\nu_e)=0.549123$, resulting in an estimated ratio of $\xi_M=(0.399498:0.300331:0.300170)$;  $\xi_{BH}=(0.316636:0.3447809:0.338583)$;
$\xi_{V}=(0.360436:0.317195:0.322369)$ where the subscripts, $M$, $BH$ and $V$ refer to the cases: magnetar, BH and the vacuum, respectively. Based on this ratio, we may determine which type of progenitor was involved during LGRB creation. For example, if we somehow detect the same 12 MeV neutrino with a percentage near to $\xi_M$, we may assume that the LGRB was produced by a magnetar. In contrast, detecting the identical neutrino with a $\xi_{BH}$ percentage confirms that the progenitor was a BH. It is worth noting that we repeated the computations using both NO and IO mass hierarchies, but the variations in our energy range were negligible. These fluctuations can be seen first in the oscillation probabilities, and subsequently in the incident neutrino flavor rates.\\

Recognizing these neutrino properties is meaningful since it is currently thought that there are a large number of sources with hidden jets that could contribute to the observed diffuse neutrino flux. Failed GRB with choked jets have been proposed as candidate sources in this context (among Low-Luminosity and Low-Power GRB) \citep{raz10,fra14hidden,car20}. Those are bursts where the jet is unable to escape the putative star's outer layers, resulting in an unobservable EM counterpart. Therefore, the medium is densely opaque to photons and transparent to neutrinos. In this way, the early escaping neutrinos  provide tremendous opportunity to characterize the progenitors of these sources that we would not otherwise  observe. We want to emphasize that since the EM signal cannot be expected in this scenario, we can not make use of the traditional X-ray light curve analysis to discriminate between both central engine models.\\

Future neutrino detectors will be able to quantify these attributes to provide us with a picture of how these particles behave as they travel through both media.{ Actually, in this work, we have estimated the number of events expected with the Hyper-K detector from a positivist perspective, assuming an ideal detector. We believe that with  $\sim 1\times10^3$ events detected in Hyper-K, we could already make a statistic to differentiate the rate of incident flavors. Which, again, could only occur in the best-case scenario.
We know that a more accurate analysis requires a more exhaustive investigation taking into account the uncertainties and detector sensitivities, which we believe is beyond the scope of this work.}\\

{It is worth mentioning that in the more distant future, there are plans to build mega detectors such as Deep-TITAND (5 Mton) \citep{bos15} and MICA (10 Mton) \citep{kis11} that will increase the fiducial volume of Hyper-K by up to 10 and 20 times, respectively. Therefore, we believe our findings will be replicated eventually.}


\section*{Declaration of competing interest}
The authors declare that they have no known competing financial interests or personal relationships that could have appeared to influence the work reported in this paper.
\section*{Acknowledgments}
We appreciate the anonymous referee's contributions, which helped to improve this manuscript. We thank D. Page and C. G. Bernal for useful discussions. GM acknowledges the financial support through the CONACyT grant 825482. NF acknowledges the financial support from UNAM-DGAPA-PAPIIT  through IN106521.

\section*{Appendix}
Under the strong magnetic field regime ($(\Omega_B\equiv eB/m_e^2 \gg 1)$, all leptons are confined to the Landau zero level ($\lambda_n^2=0$) and therefore, the electron energy is $E_{e,0}^2=(p_3^2+m_e^2)\,,$ so the  functions of the neutrino effective potential in this strong regime are
\begin{eqnarray}
F_s&=&\left[1+ \frac{m_e^2}{m^2_W}\left(\frac32+2\frac{E^2_\nu}{m^2_e} +\frac{B}{B_c}\right)\right]K_1(\sigma_l)\,, \hspace{1cm}  G_s=\left[1+ \frac{m_e^2}{m^2_W}\left(\frac12-2\frac{E^2_\nu}{m^2_e} +\frac{B}{B_c}\right)\right]K_1(\sigma_l)\,,\cr
J_s&=&\frac34K_0(\sigma_l)+\frac{K_1(\sigma_l)}{\sigma_l} \,, \hspace{3.8cm}    H_s=\frac{K_1(\sigma_l)}{\sigma_l}\,,
\end{eqnarray}

where $m_e$ is the electron mass, $m_W$ is the mass of the W boson, $\sigma_l=(l+1)m_e/T$  with $\mu$ and $T$ the chemical potential and temperature, respectively. Finally, $K_j$ represents the modified Bessel function of order $j$.\\

Under the mild magnetic field regime ($\Omega_B<1$), Landau levels begin to fill gradually and the electron energy is given by
 \begin{equation}
     E_{e,n}=\sqrt{p_3^2+m_e^2+2neB}=p_3^2+m_e^2(1+2n\Omega_B)=p_3^2+m_e^2\lambda_n^2\ ,
 \end{equation}
so the  functions of the neutrino effective potential in this mild regime  are

\begin{eqnarray}
F_m&=&\biggl(1+2\frac{E^2_\nu}{m^2_W}\biggr)K_1(\sigma_l)+2\sum_{n=1}^\infty\lambda_n\left(1+\frac{E^2_\nu}{m^2_W} \right)K_1(\sigma_l\lambda_n) \,,  \nonumber\\ G_m&=&\left(1+2\frac{E^2_\nu}{m^2_W}\right)K_1(\sigma_l)-2\sum_{n=1}^\infty \frac{\lambda_nE^2_\nu K_1(\sigma_l\lambda_n)}{m^2_W}   \,, \nonumber\\
J_m&=& \frac{3}{4}K_0(\sigma_l)+\frac{K_1(\sigma_l)}{\sigma_l} +\sum_{n=1}^\infty\lambda_n^2\left( K_0(\sigma_l\lambda)-\frac{K_0(\sigma_l\lambda)}{2\lambda_n^2}+\frac{K_1(\sigma_l\lambda)}{\sigma_l\lambda}   \right)
\,,  \nonumber\\
H_m&=& \frac{K_1(\sigma_l)}{\sigma_l}+\sum_{n=1}^\infty\lambda_n^2\left(\frac{K_1(\sigma_l\lambda)}{\sigma_l\lambda}- \frac{K_0(\sigma_l\lambda)}{2\lambda_n^2}\right) \,,
\end{eqnarray}

where $\lambda_n$ is defined as $\lambda_n\equiv\sqrt{1+2n\Omega_B} $.

\bibliographystyle{unsrt85}
\bibliography{Bib/Bib_osc}
\addcontentsline{toc}{chapter}{Bibliography}\,
\end{document}